\documentclass[pra,aps,twocolumn,superscriptaddress,10pt,showpacs]{revtex4-2}
\usepackage{amsmath,amssymb,amsfonts,bbm,graphicx,times,psfrag}
\usepackage[pdftex]{color}
\usepackage{amsmath,bm}
\usepackage[colorlinks, linkcolor=blue, citecolor=blue, urlcolor=blue, breaklinks=true]{hyperref}
\usepackage{microtype}

\usepackage{amsthm}
\usepackage{csquotes}
\usepackage{amsthm}\newcommand{\be} {\begin{equation}}
\newcommand{\ee} {\end{equation}}

\usepackage{color}
\usepackage{soul}

\usepackage{physics}

\usepackage{nicefrac, xfrac}

\begin{document}
\title{Frequency estimation by frequency jumps}
\author{Simone Cavazzoni}
\affiliation{Dipartimento FIM, Universit\`{a} di Modena e Reggio Emilia, I-41125 Modena, Italy}
\email{simone.cavazzoni@unimore.it}
\author{Berihu Teklu}
\affiliation{College of Computing and Mathematical Sciences and 
Center for Cyber-Physical Systems (C2PS), Khalifa University, 127788, 
Abu Dhabi, United Arab Emirates}
\email{berihu.gebrehiwot@ku.ac.ae}
\author{Matteo G. A. Paris}
\affiliation{Dipartimento di Fisica {\em Aldo Pontremoli}, Universit\`{a} degli Studi di Milano, 
I-20133 Milano, Italy}
\email{matteo.paris@fisica.unimi.it}
\date{\today}

\begin{abstract}
The frequency of a quantum harmonic oscillator cannot be determined through static measurement strategies on a prepared state, as the eigenstates of the system are independent of its frequency. Therefore, dynamic procedures must be employed, involving measurements taken after the system has evolved and encoded the frequency information. This paper explores the precision achievable in a protocol where a known detuning suddenly shifts the oscillator’s frequency, which then reverts to its original value after a specific time interval. Our results demonstrate that the squeezing induced by this frequency jump can effectively enhance the encoding of frequency information, significantly improving the quantum signal-to-noise ratio (QSNR) compared to standard free evolution at the same resource (energy and time) cost. The QSNR exhibits minimal dependence on the actual frequency and increases with both the magnitude of the detuning and the overall duration of the protocol. Furthermore, incorporating multiple frequency jumps into the protocol could further enhance precision, particularly for lower frequency values.
\end{abstract}
\maketitle
\section{Introduction}
Harmonic behavior is ubiquitous in physics, and the quantum harmonic oscillator (QHO) model is relevant in nearly every field of physics. 
The kinematics and dynamics of the QHO are governed by the frequency parameter, whose determination is essential to characterize the system properly.
Indeed, accurate frequency estimation is relevant for several fields, including metrology~\cite{giovannetti2006quantum, udem2002optical, descamps2023quantum, kuenstner2024quantummetrologylowfrequency, boss2017quantum, cai2021quantum} and 
quantum sensing~\cite{degen2017quantum, pirandola2018advances, montenegro2024reviewquantummetrologysensing}, spectroscopy~\cite{roos2006designer, schmitt2021optimal, lamperti2020optical}, and precision timekeeping~\cite{ludlow2015optical}. Additionally, frequency estimation finds application in quantum communication and computation~\cite{gisin2007quantum, chen2021review}. 

As a matter of fact, the eigenstates of the quantum harmonic oscillator are independent of its frequency, meaning that the frequency cannot be determined using static strategies, i.e., repeated measurements on a prepared state. Instead, the frequency must be estimated dynamically by performing measurements after the system has evolved and encoded the frequency information. Free evolution may suffice for this purpose, but one may wonder whether more efficient methods exist to encode frequency information in the evolved state, thereby improving the precision of frequency estimation—either in absolute terms or for a fixed amount of resources (energy and time). This is precisely the scope of this paper. In particular, we investigate whether suddenly detuning the oscillator’s frequency \cite{Yi88,MaRhodes89,Jan92,tibaduiza2020time} and then returning it to its original value after a specific time interval can enhance the precision and efficiency of the estimation strategy.

Our results demonstrate that the squeezing induced by a frequency jump can effectively enhance the encoding of frequency information, significantly boosting the quantum signal-to-noise ratio (QSNR) compared to standard free evolution at the same resource cost. The QSNR exhibits minimal dependence on the actual frequency and increases with both the magnitude of the detuning and the overall duration of the protocol. Furthermore, incorporating multiple frequency 
jumps into the protocol further improves precision, especially for lower values of frequency.

The paper is structured as follow. In Section \ref{sec:model} we review the theoretical description of 
harmonic systems with time-dependent frequency, with emphasis on Gaussian states, whereas in Section \ref{sec:LQE} we provide a brief introduction of the local estimation theory. In Section \ref{sec:QEF}, we illustrate our results about frequency estimation by a single frequency jump and, in Section \ref{s:comp}, we compare the bounds to precision with those achievable by free evolution with no jumps in frequency. 
In Section \ref{s:jumps} we analyze the effects of multiple jumps in frequency and in Section \ref{sec:conclusions}, we close the paper with some concluding remarks.
\section{Harmonic oscillator with time dependent frequency}
\label{sec:model}
The harmonic oscillator with $n$ successive frequency jumps can be described by specifying 
the frequency function $\omega (t)$ as a piece-wise function over $n$ intervals. Let us denote 
the natural frequency as $\omega_{0}$, and the changed frequency as $\omega_1 = \omega_0 + \delta$. {The time of the first frequency jump from $\omega_0$ to $\omega_1$ is $t=0$ and successively the system spend a time interval $t=\tau/n$ at $\omega_1$, before coming back 
to $\omega_0$. Then up to $t=T/n$ the system remains at a frequency $\omega_0$, before 
coming back to $\omega_1$ and repeating the frequency jumps cycle.
Overall the system spend a time interval $T-\tau=(1-\alpha) T$ at $\omega_0$ and $\tau=\alpha T$ 
at $\omega_1$}, where $T$ is the total time evolution and $0\leq\alpha\leq1$. The frequency 
function $\omega (t)$ can be expressed as follows:
{
\begin{equation} \label{jump}
\omega(t) = \begin{cases}
\omega_{1}&  \text{if} \;\; m \tau_n \le t \le  (m+1) \tau_n\quad \; 0 \leq m < n \\
\omega_{0}& \text{elsewhere}\,,
 \end{cases}
\end{equation}}
where $\tau_n = \alpha T/n$.
In order to compute the quantum state of the HO at an arbitrary instant $t>0$, we need to obtain the time evolution operator for this time dependent Hamiltonian. 
Following \cite{Yi88,MaRhodes89,Jan92,tibaduiza2020time}, the Hamiltonian of the HO with a time-dependent frequency as given in Eq. (\ref{jump}):
\begin{equation} \label{Ham}
H=\frac{p^2}{2}+\frac{1}{2}\bigg(\omega_{0}^2+2\omega_{0}\eta(t)\bigg)q^{2}
\end{equation}
Here, $q$ represents the position operator, $p$ is the momentum operator, $\omega_{0}$ is the initial frequency, and $\eta(t)$ is the time-dependent  function 
encoding the frequency variation defined by  {
\begin{equation}
    \eta(t)= \sum_{m} \eta_{0} \bigg[ \Theta \left( \frac{m(\tau+T)}{n} \right) -\Theta \left(\frac{m(\tau+T)+\tau}{n} \right) \bigg],
\end{equation}}
where 
\begin{equation}
\eta_{0}=\frac{\omega_{1}^2-\omega_{0}^2}{2\omega_{0}} \quad \omega_{1}=\sqrt{\omega_{0}^2+2\omega_{0}\eta_{0}}.
\end{equation}
The evolution of the system is piece-wise time-independent, where the two time-independent 
Hamiltonians are respectively given by
\begin{align}
H_{1}& =\frac{p^2}{2}+\frac{1}{2}\omega_{1}^2q^{2} \notag \\ 
& = \frac12 \eta_0 (a^2 + a^{\dag 2} ) + (\omega_0 + \eta_0) (a^\dag a +\frac12) \\
H_{0}& =\frac{p^2}{2}+\frac{1}{2}\omega_{0}^2q^{2} = \omega_0 (a^\dag a +\frac12) \,,
\end{align}
where $[a,a^\dag]=1$ are the usual field operators for the QHO. The Hamiltonian $H_0$ describes free evolution and corresponds to rotation in the phase space, whereas the Hamiltonian $H_1$ describes (generalized) squeezing. Starting from the ground state of the harmonic oscillator, the evolved
state in the case of a single jump in frequency is given by 
\begin{align}
|\psi (t) \rangle & = \left\{
\begin{array}{lc}
|\psi_s (t) \rangle& 0<t<\tau\\
& \\
|\psi_\tau (t) \rangle & \tau<t< T
\end{array}
\right. \label{psit}
\end{align}
where 
\begin{align}
|\psi_s (t) \rangle & =  N(t) \sum_{n=0}^{\infty}\,\frac{\sqrt{2n!}}{2^n n!}\,
 \Lambda^n(t)\,|2n\rangle\,, \\
|\psi_\tau (t) \rangle &=  N(\tau)\sum_{n=0}^{\infty}\,\frac{\sqrt{2n!}}{2^n n!}\,
  \Lambda^n(\tau)\, e^{-2i\omega_0 n (t-\tau)}\,  |2n\rangle \,.
\end{align}
In other words, $|\psi (t) \rangle$ is a squeezed vacuum state with time-dependent 
amplitude and phase.  In the above formula we have 
\begin{align}
N(t) & = \left|\left[ \cosh{\nu(t)}-\frac{\lambda(t)}{2\nu(t)} \sinh{\nu(t)} \right]^{-2}\right|^{\frac14} \\ 
\Lambda (t) & = \frac{-4 i\, \eta_0\, t\, \sinh{\nu(t)}}{2\nu \cosh{\nu(t)}-\lambda(t) \sinh{\nu(t)}},
\end{align}
where
\begin{align}
\lambda(t) & =-2i(\omega_0+\eta_0 t) \\
\nu(t) & = \left(\frac{1}{4}{\lambda}_{3}^{2}-\eta_0^{2} t^2\right)^{\frac12}\,.
\end{align}

In order to solve the dynamics in the case of multiple jumps, it is convenient to describe the dynamics
in the phase space \cite{f05gauss}. This is possible because both the Hamiltonians above are quadratic in the field operators with two main consequences: 1. the dynamics maintains the Gaussian character of any initial Gaussian states; 2. for an initial Gaussian state, the dynamics may be entirely described using the 
symplectic formalism, i.e., by the evolution of the first two moments of the canonical operators $q$ and
$p$, i.e., the vector of mean values $X(t)=(\langle q \rangle,\langle p \rangle)$ and 
the covariance matrix ${\boldsymbol \sigma}(t)$ with elements 
$\sigma_{lm}=\frac{1}{2}\langle X_l X_m + X_m X_l \rangle - \langle X_l \rangle \langle X_m \rangle$
where $\langle...\rangle = \langle\psi(t)|...|\psi(t)\rangle$.

In particular, if we start from the QHO initially in the ground state we have $X(0)=(0,0)$ and ${\boldsymbol \sigma}(0) = \frac12 {\mathbb I}_2$, and the state in Eq. (\ref{psit}) may be equivalently described as the Gaussian state having 
\begin{align}
X (t) & = (0,0) \\
{\boldsymbol \sigma}(t) & = \left\{\begin{array}{lc} \frac12 S(t)^T S(t) & 0<t<\tau\\ & \\ 
\frac12 R(t)^T S(\tau)^T S(\tau) R(t)& \tau<t<T\end{array}  \right.\,,
\end{align}
where ${}^T$ denotes transposition. The symplectic matrices corresponding 
to squeezing and rotation are given by 
\begin{align}
    \label{eq:rotation}
R(t) & = \cos{\omega_0 t}\, {\mathbb I} + i \sin{\omega_0 t}\, \sigma2 \\
%
%
    \label{eq:squeezing}
    S(t) & = \cosh{2r}\, {\mathbb I} + \sinh{2r}\cos{\phi}\, \sigma_3 + \sinh{2r}\sin{\phi}\, \sigma_1
\end{align}
where the $\sigma$'s are the Pauli matrices, and the time-dependent squeezing and 
phase parameters may be obtained from the relations $\hbox{Tanh}\, r(t) = |\Lambda(t)|$ and $\hbox{Tan} \phi(t) = \hbox{Arg}\,\Lambda(t)$, i.e., 
\begin{align}
    \label{eq:squeezing_par}
    r(t) & = \hbox{ArcTanh}\left( \frac{|\omega_0^2-\omega_1^2|}{\sqrt{(\omega_0 ^2 + \omega_1 ^2)^2+ 4\omega_0^2 \omega_1^2 \cot^2(\omega_1 t) }}\right) \\ 
    \phi(t) & = \hbox{ArcTan}  \left( \frac{2\omega_0 \omega_1 \cot(\omega_1 t)}{\omega_0 ^2 + \omega_1 ^2} \right)
    \label{eq:phase_squeezing_par}
\end{align}
In the case of $n$ frequency jumps, at the end of the cycle the vector of mean values still vanishes, whereas the covariance matrix is given by 
\begin{align}
{\boldsymbol \sigma}_n(\alpha,T) = \frac12 \left[ R^T(T_n)S^T(\tau_n)\right]^n  \left[S(\tau_n)R(t_n) \right]^n 
\end{align}
where
\begin{align} \label{splits}
\tau_n  = \frac{\alpha T}{n} \qquad
T_n  = \frac{(1-\alpha)T}{n}
\end{align}
\section{Local quantum estimation theory} 
\label{sec:LQE}
By encoding a parameter onto the states of a quantum system, we obtain a family 
of density matrices $\rho_\lambda$, usually referred to as a {\em quantum 
statistical model}, $\lambda\in\Lambda\subset {\mathbb R}$. An estimation strategy consists of an observable 
to be measured and an estimator to process data. The measurement is 
described by a positive operator-valued measure (POVM) 
$\left\{\Pi_y\right\}$, $y\in Y$, such that
$\Pi_y > 0\,, \forall y$, and $\sum_{y\in Y} \Pi_y = {\mathbb I}$. 
The estimator is a function $\hat\lambda$ from the data space $Y\times Y\dots \times Y$ ($M$ times) to the domain $\Lambda$, where $M$ denotes the number of repeated measurements. The outcomes of the measurement are distributed according to the Born rule $p(y|\lambda) = \hbox{Tr}\left[\rho_\lambda\, \Pi_y\right]$. The precision of the estimation strategy is quantified by the variance of the estimator. For unbiased estimators, i.e., $\int_Y dy\, p(y|\lambda)\, \hat\lambda(y) = \lambda$, the Cram\`er-Rao theorem establishes a bound on the variance as follows 
\begin{equation}\label{eq:CRB}
\text{Var} \hat \lambda\geq\frac{1}{M F(\lambda)}\, ,
\end{equation}
where the Fisher information $F(\lambda)$ is defined as 
 \begin{equation}\label{eq:FI}
F(\lambda)  = \int\!\! dy \, p(y | \lambda)\big[\partial_\lambda \ln p(y | \lambda)\big]^2\,.
\end{equation}
An estimator is said to be {\em efficient} if it saturates the Cram\`er-Rao bound.
The ultimate bound on the precision of any estimation strategy for $\lambda$ may be obtained 
by maximizing the Fisher information over all the possible POVM. The optimization may be 
actually carried out and the optimal POVM corresponds to the spectral 
measure of the {\em symmetric logarithmic derivative} $L_\lambda$, which is the selfadjoint 
operator solving the Lyapunov equation
\begin{equation}\label{SLD}
2 \partial_\lambda \rho_\lambda  = L_\lambda \rho_\lambda + \rho_\lambda L_\lambda\,.
\end{equation}
The maximum value of the Fisher information is usually referred to as the 
Quantum Fisher Information (QFU) $G(\lambda)$, and the corresponding bound as the 
Quantum Cram\`er-Rao bound
\begin{align}
\max_{\{\Pi_y\}} F(\lambda) & = G(\lambda) \equiv \hbox{Tr} \left[\rho_\lambda\, L_\lambda^2\right] \\
\text{Var} \hat \lambda& \geq\frac{1}{M G(\lambda)}\,.
\end{align}
For statistical models made of pure states $\rho_\lambda=|\psi_\lambda\rangle\langle\psi_\lambda|$ 
the QFI may be written as 
\begin{align}
\label{pureH}
G(\lambda) = 4 \left[ \langle\partial_\lambda \psi_\lambda|\partial_\lambda \psi_\lambda\rangle 
- \left|\langle\partial_\lambda \psi_\lambda |\psi_\lambda\rangle \right|^2\right]\,.
\end{align}
For Gaussian states having vanishing vector of mean values $X=(0,0)$ and covariance matrix 
${\boldsymbol \sigma}$ the QFI may be written as 
\begin{equation}
    \label{QFIg}
     G(\lambda)=- \hbox{Tr}\left[\Omega^T \left(\partial_\lambda{\boldsymbol\sigma}\right)\,\Omega\, \left(\partial_\lambda{\boldsymbol\sigma}\right)\right]\,,
\end{equation}
where $\Omega=i \sigma_2$ is the symplectic matrix.

Finally, in order to fairly compare estimation schemes for small and large actual values of 
the parameter, we introduce the signal-to-noise ratio (SNR) of an estimation strategy, 
$R(\lambda) = \lambda^2 /\hbox{Var} \lambda \leq \lambda^2 F(\lambda)$. The optimal measurement 
is characterized by the maximal value of the SNR and, in general, we have $R(\lambda) \leq Q(\lambda)$, 
where the quantum signal-to-noise ratio (QSNR) is defined by 
\begin{equation}
    \label{eq:QSNR}
    Q(\lambda) = \lambda^2 G(\lambda)\,.
\end{equation}
Local QET has been successfully applied to find the ultimate bounds to precision 
for estimation problems in open quantum systems, non-unitary processes, and nonlinear 
quantities as entanglement \cite{Chaohong2024,Chuan2020,Peng2018,Adolfo2017,Adesso2016,Macieszczak,genoni2008optimal,paris2009quantum,brida2010experimental,brivio2010experimental,pinel2013quantum,teklu2010phase,invernizzi2011optimal,genoni2011optical,genoni2012optical}
and for a closed system, evolving under a unitary transformation \cite{Delgado,Alipour2014,boixo2007generalized}.
The geometric structure of QET has been exploited to assess the quantum
criticality as a resource for quantum estimation \cite{Adani2024,Gietka2022understanding,Wald2020,Pieter2020,zanardi2008quantum}. 
\section{Frequency estimation by a single frequency jump}
\label{sec:QEF}
\begin{figure}[h!]
\includegraphics[width=0.48\columnwidth]{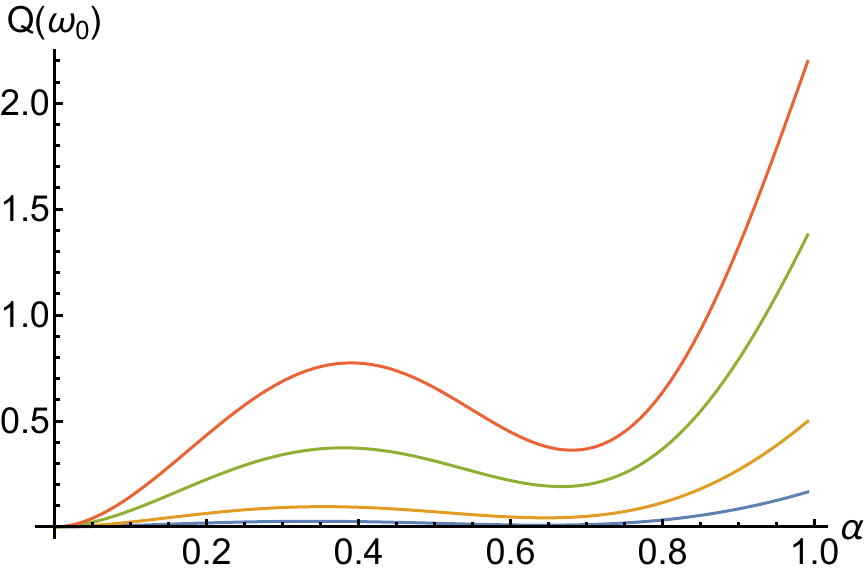}
\includegraphics[width=0.48\columnwidth]{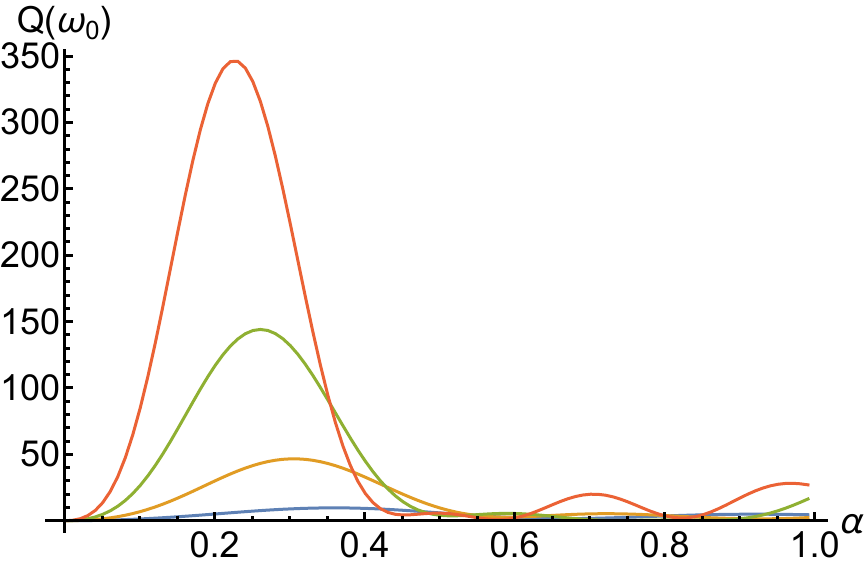}
\includegraphics[width=0.48\columnwidth]{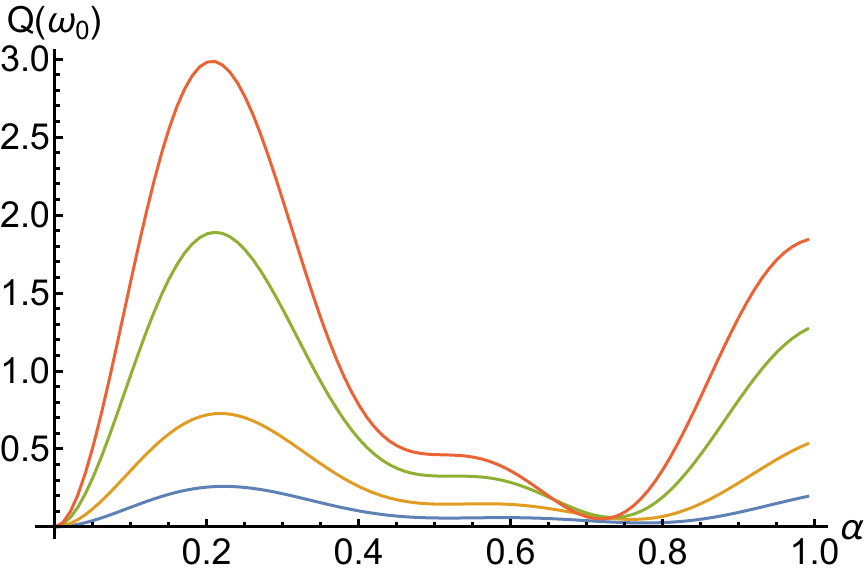}
\includegraphics[width=0.48\columnwidth]{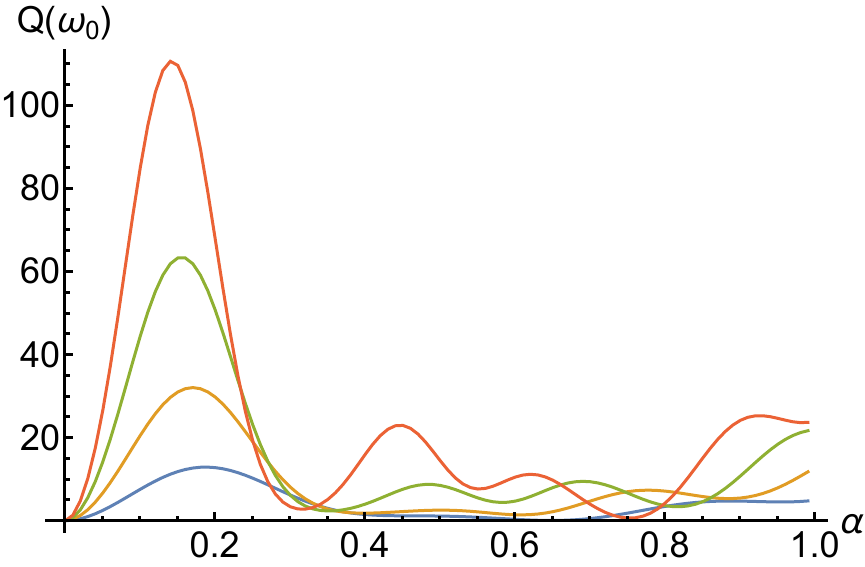}
\caption{The quantum signal-to-noise ratio $Q(\omega_0)$ as a function of the squeezing time 
fraction $\alpha$ for different values of frequency $\omega_0$ and of 
the frequency shift $\delta$. The upper plots show results for 
$\omega_0=1$ and the lower ones for $\omega_0=5$. In the left panels, 
we show (from bottom to top) the curves for $\delta=0.3$ (blue), 
$\delta=0.5$ (yellow), $\delta=0.8$ (green), $\delta=1.9$ (orange), respectively. 
In the right panels we show (from bottom to top) the curves for 
$\delta=2$ (blue), $\delta=3$ (yellow), $\delta=4$ (green), 
$\delta=5$ (orange), respectively. \label{f:singleJ}}
\end{figure}

In this Section, we analyze the behavior of the quantum signal-to-noise ratio 
of the quantum Fisher information for the frequency of the oscillator, as obtained 
by performing measurements on 
the evolved state after a single frequency jump. Specifically, we optimize the duration 
of the jump, $\tau = \alpha T$, which represents the fraction of the 
total evolution time $T$ that induces squeezing, to maximize the QSNR and the QFI, thus 
enhancing the metrological  properties of the probe. In the following, we continue 
to denote by $\omega_0$ the natural frequency of the oscillator, i.e., the parameter 
to be estimated, and introduce the symbol $\delta = \omega_1 - \omega_0$ to represent 
the frequency shift. At first, we set $T=1$ and defer the discussion about the dependence 
on $T$ to the end of the Section. Results are summarized in the four panels of Fig. 
\ref{f:singleJ}. 
\par\noindent 
For lower values of the frequency and frequency shift, i.e., $\omega_0 \lesssim 1$ and $\omega_0 \gtrsim \delta$, the optimal procedure is to avoid free evolution. This means choosing $\alpha = 1$ and dedicating the entire evolution time to the squeezing process. This is illustrated in the upper left panel of Fig. \ref{f:singleJ}, where we also observe that the QSNR  is not monotonic with respect to $\alpha$ but instead exhibits a local maximum. Additionally, we note that the QSNR increases rapidly with $\delta$.  As the frequency shift increases to larger values, $\delta \gtrsim \omega_0$, the local maximum of the QSNR surpasses the value for $\alpha = 1$, and a non-trivial optimal value for the jump duration emerges, as shown in the upper right panel of Fig. \ref{f:singleJ}.

A similar behavior, namely the appearance of a non-trivial optimal value of $\alpha$, is observed for larger values of the frequency. Results for $\omega_0 = 5$ are displayed in the lower left panel of Fig. \ref{f:singleJ}. This trend is further confirmed for larger $\delta$, as seen in the lower right panel of the same figure. The value of the QSNR is not significantly affected by the specific value of the frequency, indicating that the QFI decreases approximately as $G(\omega_0) \propto \omega_0^{-2}$ with increasing frequency. As $\delta$ is further increased, additional peaks appear, although the first maximum, occurring at smaller $\alpha$, remains the absolute maximum.

Using Eqs. (\ref{pureH}) or (\ref{QFIg}), the analytic expression for $Q(\omega_0)$ can be derived. However, this expression is cumbersome and is not provided here. On the other hand, it is straightforward to show that the amplitude and phase of the squeezing depend on the relevant quantities through the dimensionless combinations $\omega_0\tau = \alpha\,\omega_0\, T$ and $\delta/\omega_0$, while the QFI and QSNR depend on $\alpha\, T$ and $\delta/\omega_0$. This implies that the optimal time fraction 
maximizing the squeezing amplitude, i.e., $
\alpha_{\text{max}} = \frac{\pi}{2}[T(\omega_0 + \delta)]^{-1}$,
corresponding to $r_{\text{max}} = \log(1 + \delta/\omega_0)$,
is not in general the same as the one maximizing the QSNR, which must 
be determined numerically.

\begin{figure}[h!]
\includegraphics[width=0.48\columnwidth]{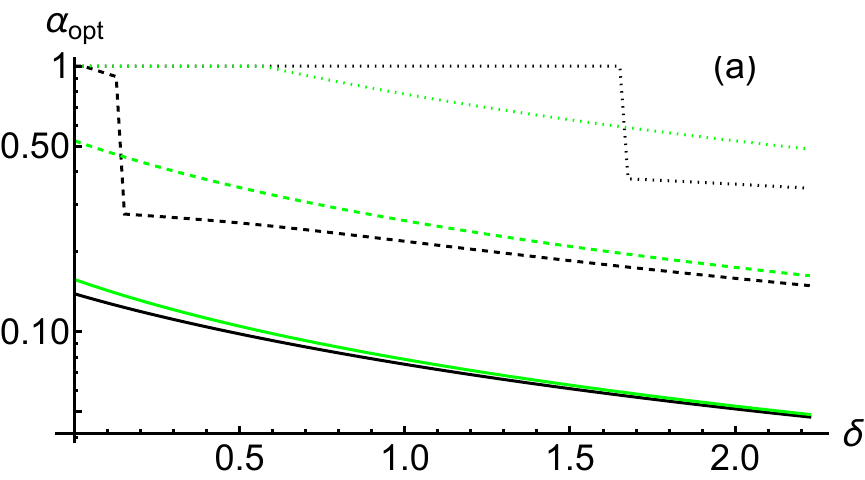}
\includegraphics[width=0.48\columnwidth]{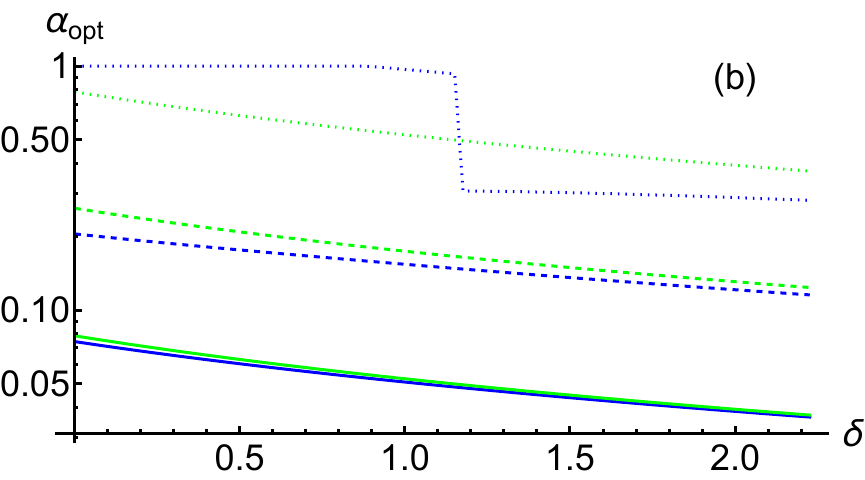}
\includegraphics[width=0.48\columnwidth]{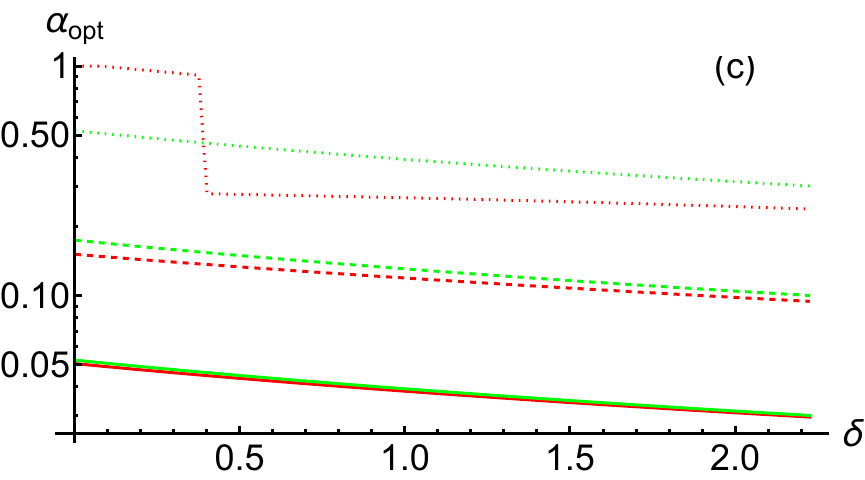}
\includegraphics[width=0.48\columnwidth]{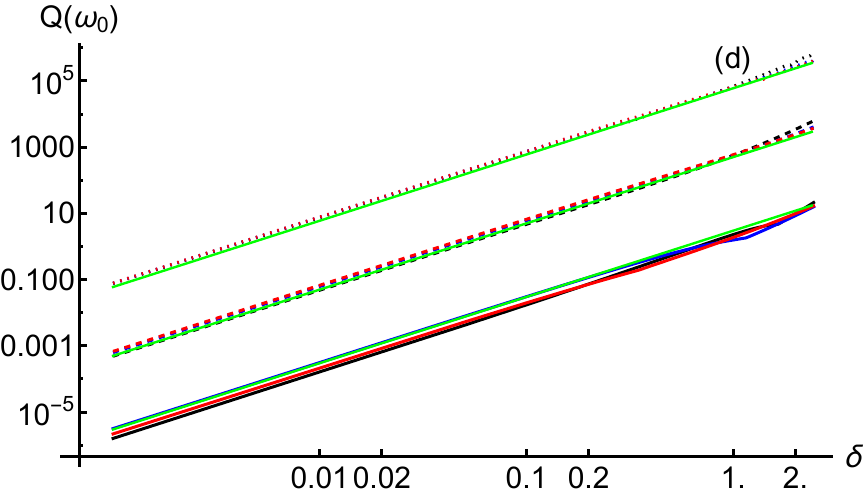}
\caption{Panels (a), (b), and (c): the optimal time fraction $\alpha_{\hbox{\footnotesize opt}}$ 
maximizing the QSNR as a function of the frequency shift $\delta$ for $\omega_0=1, 2, 3$ 
respectively. In each plot, we show results for $T=1$ (dotted line), $T=3$ (dashed), 
and $T=10$ (solid). The green lines denote the corresponding values of 
$\alpha_{\hbox{\footnotesize max}}$, the time fraction maximizing the squeezing amplitude.
Panel (d): the maximized QSNR  
as a function of the frequency shift for $\omega_0 = 1, 2, 3$ and $T = 1, 10, 100$. The curves for different values of $\omega_0$ nearly overlap for a given value of $T$. The green lines serve as visual guides proportional to $T^2 \delta^2$.
\label{f:opt}}
\end{figure}
\par\noindent
In panels (a), (b), and (c) of Fig. \ref{f:opt}, we show the optimal time fraction 
$\alpha_{\hbox{\footnotesize opt}}$ as a function of the frequency shift $\delta$ for 
different values of the frequency $\omega_0$ ($\omega_0=1,2,3$, respectively) and the 
total evolution time $T$ (from top to bottom $T=1, 3, 10$ in each plot). In the three 
panels, the green lines denote $\alpha_{\hbox{\footnotesize max}}$, i.e., the time 
fraction maximizing the amplitude of squeezing. We see that the two values may be 
very different for lower $T$, whereas they become
very close to each other for increasing $T$. The optimal time fraction 
$\alpha_{\hbox{\footnotesize opt}}$ decreases with $\delta$, consistently 
with the results shown in Fig. \ref{f:singleJ}. 

In panel (d) of Fig.~\ref{f:opt}, we show the maximized quantum SNR $Q(\omega_0)$, 
obtained for $\alpha = \alpha_{\text{opt}}$, as a function of the frequency 
shift for $\omega_0 = 1, 2, 3$ and $T = 1, 10, 100$. As evident from the plot, 
the dependence on the frequency is very weak, and the curves for different 
values of $\omega_0$ nearly overlap for a given value of $T$. The green lines 
serve as visual guides, exhibiting a behavior proportional to $T^2 \delta^2$.
\section{Comparison with free evolution}\label{s:comp}
In principle, the frequency can be estimated by performing measurements after free 
evolution (and no frequency jumps) of an initially prepared state $|\psi_0\rangle$. 
In this case, the family of states encoding frequency information is given by 
$|\psi(t)\rangle = e^{- i t \omega_0 n} |\psi_0\rangle$, with $n = a^\dag a$ 
and the QFI and QSNR may be easily evaluated as
\begin{align}
G_{\rm f} = 4 t^2 \Delta n^2 \qquad Q_{\rm f} = 4 \omega_0^2 t^2 \Delta n^2
\label{GfQf}
\end{align} 
where  $\Delta n^2 = \langle\psi_0 | n^2 
|\psi_0\rangle -  \langle\psi_0 | n |\psi_0\rangle^2$. The QFI is independent 
of the frequency, scales quadratically with the overall evolution time, 
and depends on the fluctuations of the number operator in the initial state. 
For the oscillator initially prepared in a coherent state $\Delta n^2 = 
\bar n = \langle a^\dag a \rangle$.

Estimation strategies should be compared under a fixed amount of resources.
In the case of frequency estimation, these resources are the energy of the probe 
state and the evolution time. For the free evolution, we consider the oscillator 
initially prepared in a coherent state, allowing us to safely assume that the 
time required to prepare the initial state is negligible. This enables a fair 
comparison of the two encoding strategies, both of which has a duration $T$. 
Regarding energy, in the jump-based strategy, it corresponds to the mean number 
of squeezing quanta generated during the jump, which can be evaluated using Eq. 
(\ref{eq:squeezing_par})
\begin{align}
\bar n  =& \sinh^2 r (\alpha_{\hbox{\footnotesize opt}} T) = 
\left(\frac{\delta}{2\omega_0}\right)^2
\frac{2+\frac{\delta}{\omega_0}}{1+\frac{\delta}{\omega_0}} \notag \\
&\times 
\sin^2 \left[\omega_0 \alpha_{\hbox{\footnotesize opt}} T 
\left(1+\frac{\delta}{\omega_0}\right)\right]\,.
\label{barn}
\end{align}
For the free evolution, the dynamics is passive, i.e. no energy is added, and the
number of quanta is that of the initial coherent state. 

In order to compare the two strategies, we introduce the ratio $\gamma$ between 
the QSNRs (which is equal to the ratio of the QFIs), and evaluate it for the 
oscillator initially prepared in a coherent state. According to Eq. (\ref{GfQf}) 
we have
\begin{align}
\gamma & = \frac{Q(\omega_0)}{4 T^2\, \bar n}
\label{gammas}
\end{align}
where one has to insert the expression of $\bar n$ in Eq. (\ref{barn}) in order to compare 
the two strategies using the same amount of resources. 

\begin{figure}[h!]
\includegraphics[width=0.48\columnwidth]{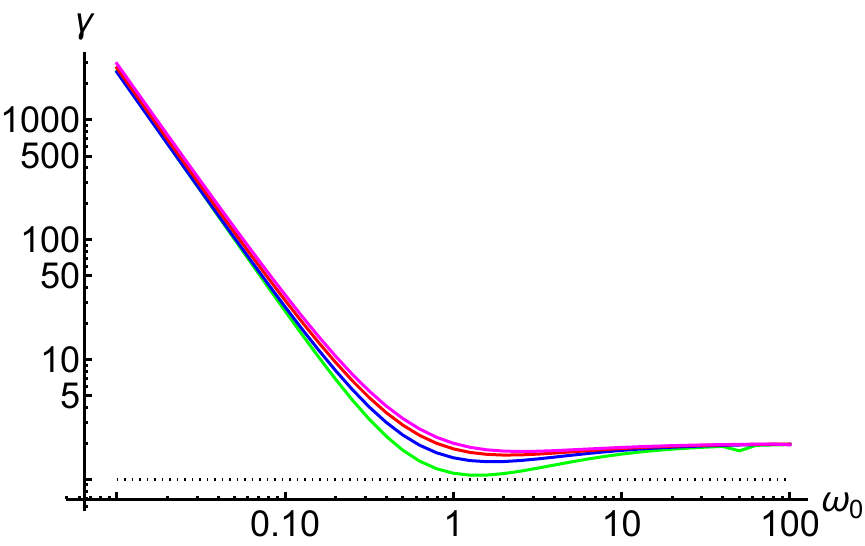}
\includegraphics[width=0.48\columnwidth]{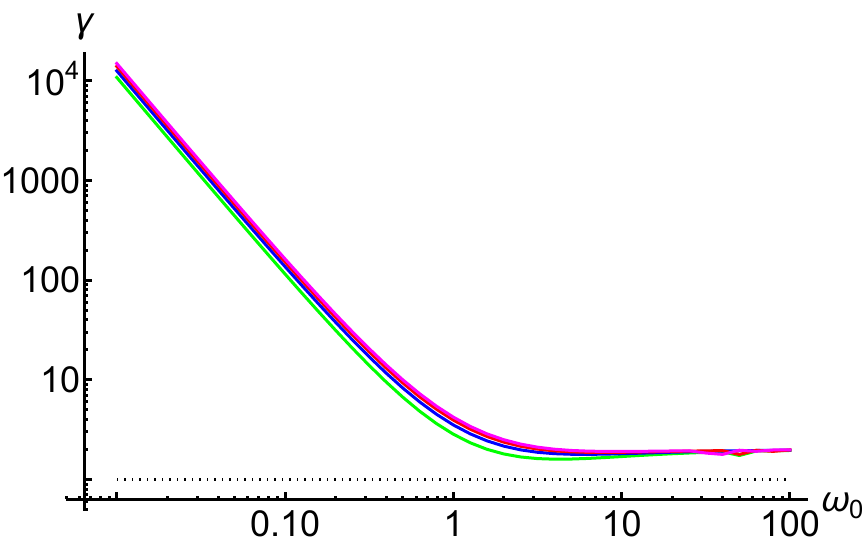}
\caption{The ratio $\gamma$  between the QSNRs 
of jump-based and free evolutions as a function of the actual value of the 
frequency $\omega_0$ for different values of the overall time duration 
(from top to bottom, $T=5, 4, 3, 2$, respectively). The left panel shows results 
for $\delta=1.0$ and the right one for  $\delta=2.0$. The asymptotic value 
for $\omega_0 \gg1$ is $\gamma\simeq 2$, independently on $T$. The dotted line 
denotes the value $\gamma=1$.
\label{f:comp}}
\end{figure}
\noindent 
In Fig. \ref{f:comp} we show the ratio $\gamma$ as a function of the 
frequency $\omega_0$ for different values of the overall time duration 
$T$ and frequency shift $\delta$. As it is apparent from  the plot, 
the jump-based strategy outperforms free evolution for any value of 
$\omega_0$, largely improving the QSNR for lower frequency. The asymptotic 
value for $\omega_0\gg 1$ is $\gamma\simeq 2$. The ratio is nearly independent 
on the duration $T$ of protocol (at least in this range of values) and, 
for lower values of $\omega_0$,  increases with $\delta$. We conclude that the 
squeezing induced by this frequency jump can effectively enhance the encoding 
of frequency information, significantly boosting the QSNR compared to free 
evolution at the same resource cost.
\section{Frequency estimation by multiple frequency jumps}
\label{s:jumps}
In Section \ref{sec:QEF}, we demonstrated that the optimal duration of the frequency 
jump is typically shorter than the total protocol duration $T$. This finding highlights 
the positive interplay between frequency jumps and free evolution. A natural 
question arises: could introducing additional jumps, interspersed with 
intervals of free evolution, further enhance frequency encoding? In this 
section, we prove that this is indeed the case. Specifically, we show 
that multiple jumps can significantly improve precision, particularly 
for lower frequency values.

\begin{figure}[h!]
\includegraphics[width=0.48\columnwidth]{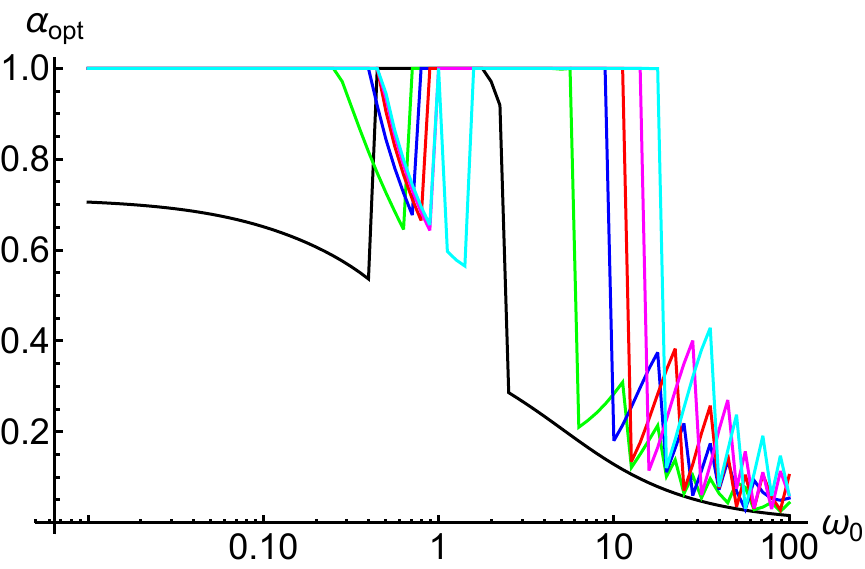}
\includegraphics[width=0.48\columnwidth]{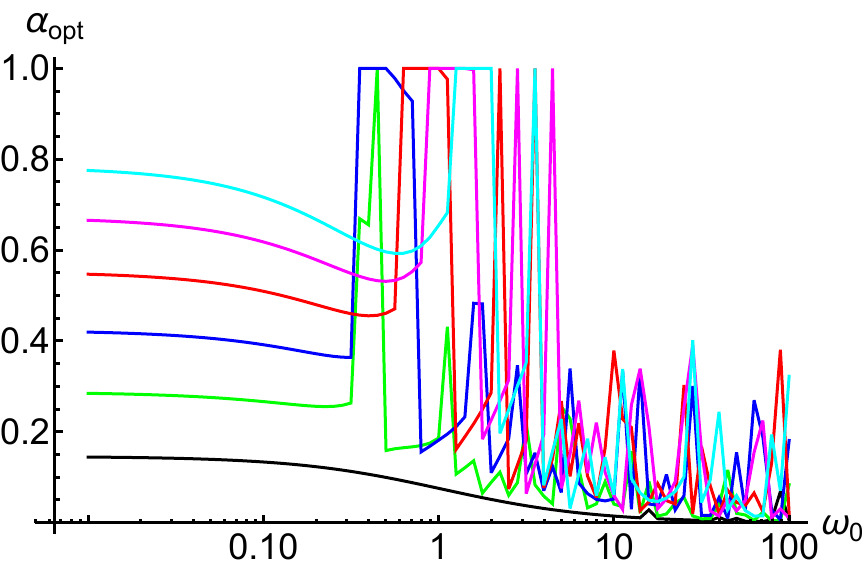}
\includegraphics[width=0.48\columnwidth]{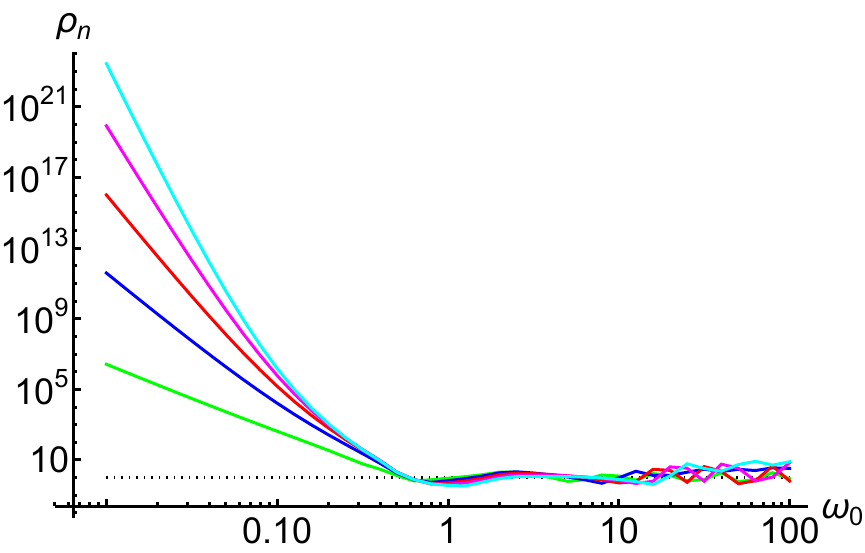}
\includegraphics[width=0.48\columnwidth]{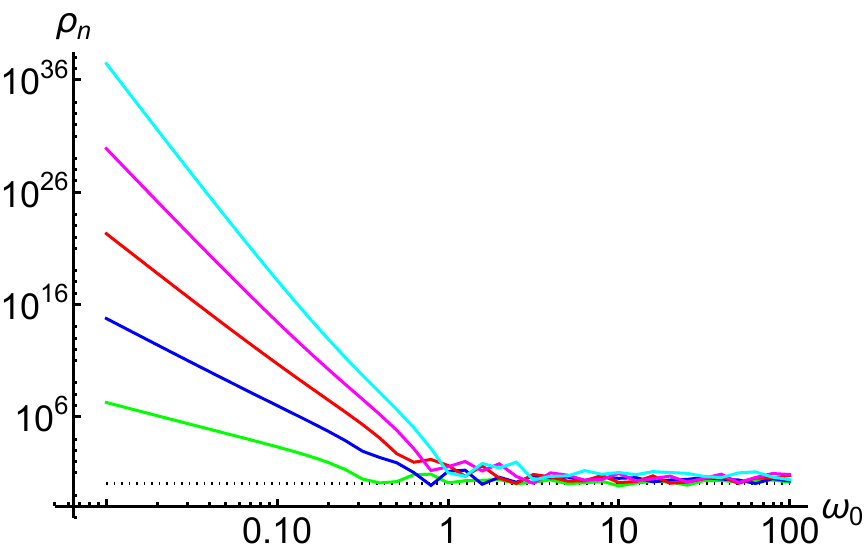}
\caption{The upper panels show the optimal values 
$\alpha_{\rm opt}$  as a function of the frequency 
for different number $n$ of jumps: $n=1$ (black),
$n=2$  (green), $n=3$  (blue), $n=4$  (red),
$n=5$  (magenta), $n=6$  (cyan), 
$\delta=1$ and two different values of the total duration $T$ of the evolution 
($T=1$ on the left and $T=10$ on the right, respectively). 
The lower panels show the corresponding ratios  $\rho_n$, defined in Eq. (\ref{rhos}), illustrating that using multiple frequency jumps consistently improves precision, with 
the enhancement being particularly pronounced for lower frequencies.
\label{f:njump}}
\end{figure}

To this end, we introduce the ratios 
\begin{align}\label{rhos}
\rho_n = \frac{Q_n(\omega_0)}{Q(\omega_0)}
\end{align}
of the QSNR obtained by dividing the total duration $T$ of the evolution into $n$
cycles, each consisting of a frequency jump followed by free evolution, as 
described in Eq. (\ref{splits}), and the corresponding QSNR obtained from 
a single jump. Note that the optimization of the jump duration $\alpha$ yields 
different values for different values of $n$. In the upper panels of Fig. 
\ref{f:njump} we show the optimal values $\alpha_{\rm opt}$ as a function 
of the frequency for different number $n$ of jumps, $\delta=1$ and two 
different values of the total duration $T$ of the evolution.
In both cases, the optimal jump duration depends on the number of jumps, 
although this dependence is not significant. In the lower panels, 
we present the corresponding ratios $\rho_n$. As seen in the plots, 
using multiple frequency jumps consistently improves precision, with 
the enhancement being particularly pronounced for lower frequency values.

\section{Conclusions} 
\label{sec:conclusions}
In this paper, we have addressed the estimation of frequency of a harmonic 
oscillator by protocols where a known detuning suddenly shifts the 
oscillator’s frequency, which then returns to its original value after a 
specific time interval. The squeezing induced by the frequency jump  
provides a metrologically effective encoding of frequency, which
enhances precision and increases the quantum Fisher information of
the resulting statistical model. In turn, the quantum signal-to-noise 
ratio increases compared to standard free evolution at the 
same resource cost, i.e., using the same amount of time and energy. 
The QSNR shows minimal dependence on the actual frequency and increases 
with both the magnitude of the detuning and the duration of the protocol. 
We have also found that by employing multiple frequency jumps, the 
estimation precision is further enhanced, in particular for lower values 
of the frequency.

Squeezing by frequency jumps has been realized experimentally in levitated 
optomechanical systems \cite{optm1,optm2} and to create squeezed states of 
atomic motion \cite{atm1}. In those
systems, the protocol presented in this paper may be implemented with current 
technology. More generally, our results pave the way to more effective encoding 
of frequency, including nonlinear squeezing and information scrambling.
\section*{Acknowledgments}
\noindent This work has been supported by Khalifa University of Science and 
Technology through the project C2PS-8474000137, and partially supported by EU 
and MIUR through the project PRIN22-2022T25TR3-RISQUE. The authors thank Paolo 
Bordone for valuable discussions and Marco Adani for his contribution in the 
early stage of this project. 

\bibliography{biblio.bib}

\end{document}